\documentclass[aps, pra,twocolumn,superscriptaddress,longbibliography]{revtex4-2}
\usepackage{amsmath,amssymb, bm}
\usepackage{booktabs}
\usepackage{graphicx}
\usepackage{braket} 
\usepackage{xcolor}
\usepackage{tikz}
\usepackage{tikz-3dplot}
\usepackage{float}
\usepackage{subfig}
\usepackage{microtype}
\usepackage{ragged2e}
\newcounter{definition}
\newenvironment{definition}
{\par\vspace{0.5em}\noindent\refstepcounter{definition}\textbf{Definition \thedefinition.}\quad}
{\par\vspace{0.5em}}
\newcounter{lemma}
\newenvironment{lemma}
{\par\vspace{0.5em}\noindent\refstepcounter{lemma}\textbf{Lemma \thelemma.}\quad}
{\par\vspace{0.5em}}
\newenvironment{proof}
{\par\vspace{0.3em}\noindent\textbf{Proof.}\quad}
{\par\hfill$\blacksquare$\par\vspace{0.3em}}
\newcounter{theorem}
\newenvironment{theorem}
{\par\vspace{0.5em}\noindent\refstepcounter{theorem}\textbf{Theorem \thetheorem.}\quad}
{\par\vspace{0.5em}}
\newcounter{myfigurecount}
\usepackage[
colorlinks=true,
citecolor=blue,       
linkcolor=blue,      
urlcolor=blue,        
backref=false
]{hyperref}

\begin{document}
	\sloppy
	\title{Local distinguishability of five six-partite orthogonal product states}
    
    \author{Ben-Hao \surname{Zhao}}%
    \affiliation{College of Computer Science and Engineering, Shandong University of Science and Technology, Qingdao, 266590, China}
   
    \author{Dong-Huan \surname{Jiang}}
    \affiliation{College of Mathematics and Systems Science, Shandong University of Science and Technology, Qingdao, 266590, China}
   
   \author{Yu-Guang \surname{Yang}}%
   \affiliation{College of Cyberspace Science and Technology, Beijing University of Technology, Beijing, 100124, China}

   \author{Guang-Bao  \surname{Xu}}%
   \email{xu\_guangbao@163.com}
   \affiliation{College of Computer Science and Engineering, Shandong University of Science and Technology, Qingdao, 266590, China}

	\date{\today}
	
	\begin{abstract}
  The local distinguishability of a set of orthogonal quantum states can be used to reduce the amount of quantum state transmission in quantum protocols and lower system operational costs and quantum resource overhead. In this paper, we investigate the local distinguishability of five six-partite orthogonal product states (OPSs), where any two states are orthogonal on only one subsystem. We classify all feasible sets of five six-partite OPSs into five categories with 249 cases in total, and prove that these sets can be perfectly distinguished by local operations and classical communication, except for two cases. For five six-partite OPSs corresponding to these two exceptional cases, we further discuss their local distinguishability. Our work fully characterizes the local distinguishability of all feasible sets of five six-partite OPSs.
  \end{abstract}
	
	\maketitle
	
	\section{Introduction}
	Local distinguishability of orthogonal quantum states, which is a current research hotspot in quantum information theory, has attracted extensive academic attention since its proposal. Initially, people intuitively believed that a set of orthogonal product states (OPSs) could be distinguished. However, in 1999, Bennett et al. first constructed two sets of OPSs, both of which cannot be perfectly distinguished by local operations and classical communication (LOCC) \cite{PhysRevA.59.1070}. This phenomenon is known as quantum nonlocality without entanglement (QNWE). The work not only subverted people’s previous cognition, but also is of pioneering significance for the field of local distinguishability of orthogonal quantum states.
	
	Since Bennett put forward the concept of QNWE, a large number of scholars have devoted themselves to the research in this field. In 2000, Walgate et al.\ first proved that any set containing two orthogonal pure states can be exactly distinguished by LOCC~\cite{PhysRevLett.85.4972}. Subsequently in 2002, Walgate et al.\ gave a complete characterization for the two‑qubit $\mathbb{C}^2\otimes\mathbb{C}^2$ Hilbert space and drew two conclusions: the condition for three orthogonal states to be perfectly distinguishable is that at least two of them are product states; the condition for four orthogonal states in this system to be perfectly distinguishable is that all of them are product states~\cite{PhysRevLett.89.147901}. DiVincenzo et al.~\cite{DiVincenzo2003} further revealed the quantitative threshold law for the local distinguishability of OPSs under different systems. For general multipartite quantum systems, as long as the total number of OPSs is no more than 3, the set is definitely perfectly distinguishable under LOCC. For bipartite quantum systems, this threshold can be relaxed to 4; That is, any set of bipartite OPSs with no more than 4 elements can be perfectly distinguished by  LOCC. The above research results provide a theoretical foundation for the problem of local distinguishability of quantum systems.
	
	Research on the local distinguishability of quantum state set has developed over many years and accumulated abundant theoretical results. In the past decade, research on constructing locally indistinguishable set of orthogonal quantum states has gradually become a core hot topic in this field. The relevant research system has gradually extended from the original bipartite quantum systems to multipartite quantum systems, and various novel and efficient construction methods have been successively proposed and refined by the academic community~\cite{PhysRevLett.87.277902,PhysRevLett.90.047902,PhysRevLett.92.177905,PhysRevLett.98.230502,PhysRevA.90.022313,PhysRevA.92.012332,PhysRevA.92.032313,Xu2016,PhysRevA.93.012314,Wang2016,Li2019,PhysRevA.102.032211,Zhang2025,PhysRevA.110.022427,PhysRevA.109.022428,PhysRevA.104.052433,PhysRevA.108.012418,PhysRevA.111.052434,Zhu2022}. With the continuous deepening of research on the local distinguishability of quantum state set, Halder et al.~\cite{PhysRevLett.122.040403} innovatively put forward the core concept of strong nonlocality. On this basis, the research on constructing sets of OPSs and orthogonal entangled states with strong quantum nonlocal features has rapidly become a popular research direction in the field and attracted extensive attention~\cite{PhysRevA.104.012424,PhysRevA.102.042202,PhysRevA.102.042228,PhysRevA.107.042214,PhysRevA.109.052422,PhysRevA.105.022209}.
	
	Although many important research achievements have been obtained in the field of quantum nonlocality as mentioned above, there are still a large number of fundamental problems that have not been effectively solved. The problem of local distinguishability in multipartite quantum systems has not been completely resolved. Shortly after Xu et al.~\cite{ht8g-43kp} solved the local distinguishability problem of four OPSs in multipartite quantum systems, they addressed the local distinguishability of five OPSs in bipartite and tripartite quantum systems~\cite{53hk-wkd3}. However, the local distinguishability problem of five OPSs in six‑partite quantum systems remains unsolved. Compared with previous studies, the case of six‑partite quantum systems is more complicated. This paper investigates the local distinguishability of five six-partite OPSs, where any two states within this set are orthogonal only on one subsystem, and proves that, except for two special cases, all remaining cases can be perfectly distinguished by LOCC.

	The paper is organized as follows. We present some preliminary knowledge in Section ~\ref{sec:2}. Section ~\ref{sec:3} discusses all cases that can be perfectly distinguished by LOCC. In Section ~\ref{sec:4}, we further investigate the distinguishability of the remaining two special cases. We conclude this paper with final remarks in Section ~\ref{sec:5}.

	\section{PRELIMINARIES}
	\label{sec:2}
	In this section, some preliminaries, which will be used in what follows, are introduced.
	
	\begin{definition}\label{def:undirected_graph}\cite{bondy2008graph}
		(Undirected graph) An undirected graph $G$ is a pair $(V, E)$ where $V$ is a finite, nonempty set of vertices and $E$ is a collection of unordered pairs of distinct vertices, called edges. For an edge $e = \{u, v\} \in E$, we say that $e$ joins the vertices $u$ and $v$; equivalently, $u$ and $v$ are incident with $e$.
	\end{definition}
	
	\begin{definition}\label{def:simple_graph}\cite{bondy2008graph}
		(Simple graph) A simple graph is an undirected graph with neither loops nor multiple edges.
	\end{definition}
	
	\begin{definition}\label{def:orthogonality_graph}\cite{DiVincenzo2003}
		Let $H=\bigotimes_{i=1}^{m} H_i$ be a $m$-partite Hilbert space with $\dim H_i = d_i$.
		Let $S=\{|\psi_j\rangle\equiv \bigotimes_{i=1}^{m}|\phi_{i,j}\rangle \,\big|\, j=1,2,\dots,n\}$ be an orthogonal product basis in $H$.
		We represent $S$ as a graph $G=(V, E_1\cup E_2\cup\cdots\cup E_m)$, where the set of edges $E_i$ have color $i$.
		The states $|\psi_j\rangle\in S$ are represented as the vertices $V$.
		There exists an edge $e$ of color $i$ between the vertices $v_k$ and $v_l$, i.e., $e\in E_i$, when states $|\psi_k\rangle$ and $|\psi_l\rangle$ are orthogonal on $H_i$.
		Since all the states in the product basis are mutually orthogonal, every vertex is connected to all the other vertices by at least one edge of some color.
		The graph $G$ is called the orthogonality graph of the product basis.
	\end{definition}
	
	\begin{definition}\label{def:bipartite_ortho_vector}\cite{53hk-wkd3}
		(The vector of the numbers of pairwise orthogonal relations) A vector $(a,b)$ is used to represent the numbers of pairwise orthogonal relations for bipartite OPSs, where $a$ denotes the number of pairwise orthogonal relations among those states on the first subsystem while $b$ denotes the number of pairwise orthogonal relations on the second subsystem. For simplicity, we call $(a,b)$ the vector of the numbers of pairwise orthogonal relations.
	\end{definition}
	
	\begin{lemma}\label{lem:1}\cite{DiVincenzo2003}
		Let $S$ be a set of bipartite OPSs with four or fewer members in any dimension (that allows for this set of OPSs). The set $S$ is distinguishable by local incomplete von Neumann measurements and classical communication.
	\end{lemma}
	
	\begin{lemma}\label{lem:2}\cite{DiVincenzo2003}
		A set of any multipartite OPSs with three or fewer members is distinguishable by local incomplete von Neumann measurements and classical communication.
	\end{lemma}
	
	It should be pointed out that a von Neumann measurement is a special kind of positive operator-valued measure (POVM).
	
	By the vectors of the numbers of pairwise orthogonal relations, all feasible sets of four tripartite OPSs can be classified into three categories, i.e., (4, 1, 1), (3, 2, 1), and (2, 2, 2) \cite{ht8g-43kp}. Their local distinguishability is presented in the lemmas below.
	
	\begin{lemma}\label{lem:3}\cite{ht8g-43kp}
		Four tripartite OPSs with the vector of the numbers of pairwise orthogonal relations $(4,1,1)$ or $(3,2,1)$ can be perfectly distinguished by LOCC.
	\end{lemma}
	
\begin{figure}[H] 
		\centering
		\includegraphics[width=0.8\linewidth]{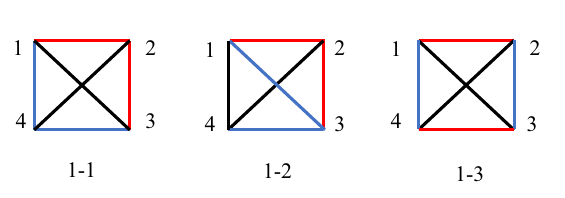}
		\caption{The feasible orthogonality graphs of four tripartite OPSs with the vector of the numbers of pairwise orthogonal relations $(2, 2, 2)$.}
		\label{fig:1}
	\end{figure}

	For category $(2,2,2)$, there exist three different orthogonality graphs, i.e., cases (1-1), (1-2), and (1-3), as shown in Fig.~\ref{fig:1}.
	
	\begin{lemma}\label{lem:4}\cite{ht8g-43kp}
		Four tripartite OPSs with orthogonality graph (1-1) or (1-2) can be perfectly distinguished by LOCC when their vector of the numbers of pairwise orthogonal relations is $(2,\,2,\,2)$.
	\end{lemma}
	
	\begin{lemma}\label{lem:5}\cite{ht8g-43kp}
		Four tripartite OPSs with orthogonality graph (1-3) cannot be perfectly distinguished by LOCC when the local ranks of the four tripartite OPSs satisfy $r_1=r_2=r_3=2$; four tripartite OPSs with orthogonality graph (1-3) can be distinguished by LOCC with some certain probability for any other case.
	\end{lemma}
	
	\begin{lemma}\label{lem:6}\cite{ht8g-43kp}
		On a four-partite quantum system, any four OPSs in any dimension (that allows for these OPSs) can be perfectly distinguished by LOCC.
	\end{lemma}
	
	\begin{lemma}\label{lem:7}\cite{ht8g-43kp}
		On a five-partite quantum system, any four OPSs in any dimension (that allows for these OPSs) can be perfectly distinguished by LOCC.
	\end{lemma}
	
	\begin{figure}[H]
		\centering
		\includegraphics[width=0.35\linewidth]{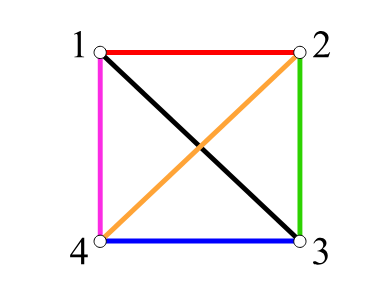}
		\caption{The possible orthogonality graph of four OPSs with the vector of the numbers of pairwise orthogonal relations (1, 1, 1, 1, 1, 1).}
		\label{fig:2}
	\end{figure}
	
	By default, in all orthogonality graphs in this paper, black edges denote orthogonality of two OPSs on only the first subsystem, red edges on only the second subsystem, blue edges on only the third subsystem, green edges on only the fourth subsystem, pink edges on only the fifth subsystem, and yellow edges on only the sixth subsystem.
	
	\begin{lemma}\label{lem:8}\cite{ht8g-43kp}
		On a six-partite quantum system, any four OPSs in any dimension (that allows for these OPSs) can be perfectly distinguished by LOCC (as shown in Fig.~\ref{fig:2}).
	\end{lemma}
	
	\section{Local Distinguishability of Five Orthogonal Product States on Six-partite Quantum Systems}
	\label{sec:3}
	In this section, we analyze the distinguishability of five six-partite OPSs, where any two states are orthogonal on only one subsystem. Clearly, the local distinguishability of five six‑partite OPSs characterized by the vector of the numbers of pairwise orthogonal relations $(a,b,c,d,e,f)$ is invariant under the interchange of subsystems. Therefore, it suffices to consider only one representative case of the vector $(a,b,c,d,e,f)$ for five six‑partite OPSs.
	
\begin{widetext}
			\begin{figure*}[tbp]
				\centering
				\includegraphics[width=0.8\linewidth]{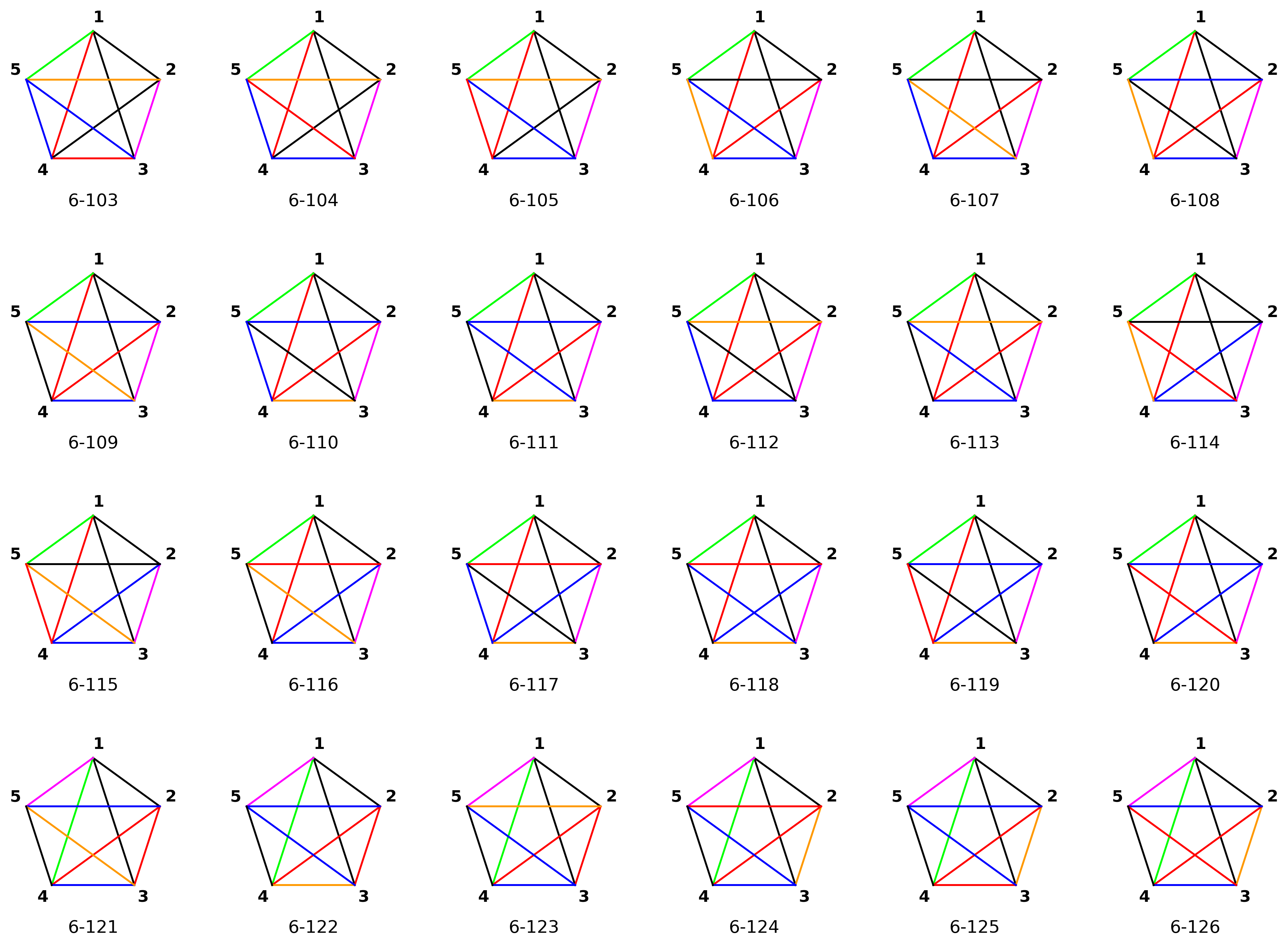}
				\caption{The feasible graphs of five six‑partite OPSs with the vector of the numbers of pairwise orthogonal relations $(3,2,2,1,1,1)$.}
				\label{fig:6}
			\end{figure*}
		\end{widetext}

\begin{figure*}[tbp]
			\centering
			\includegraphics[width=0.75\linewidth]{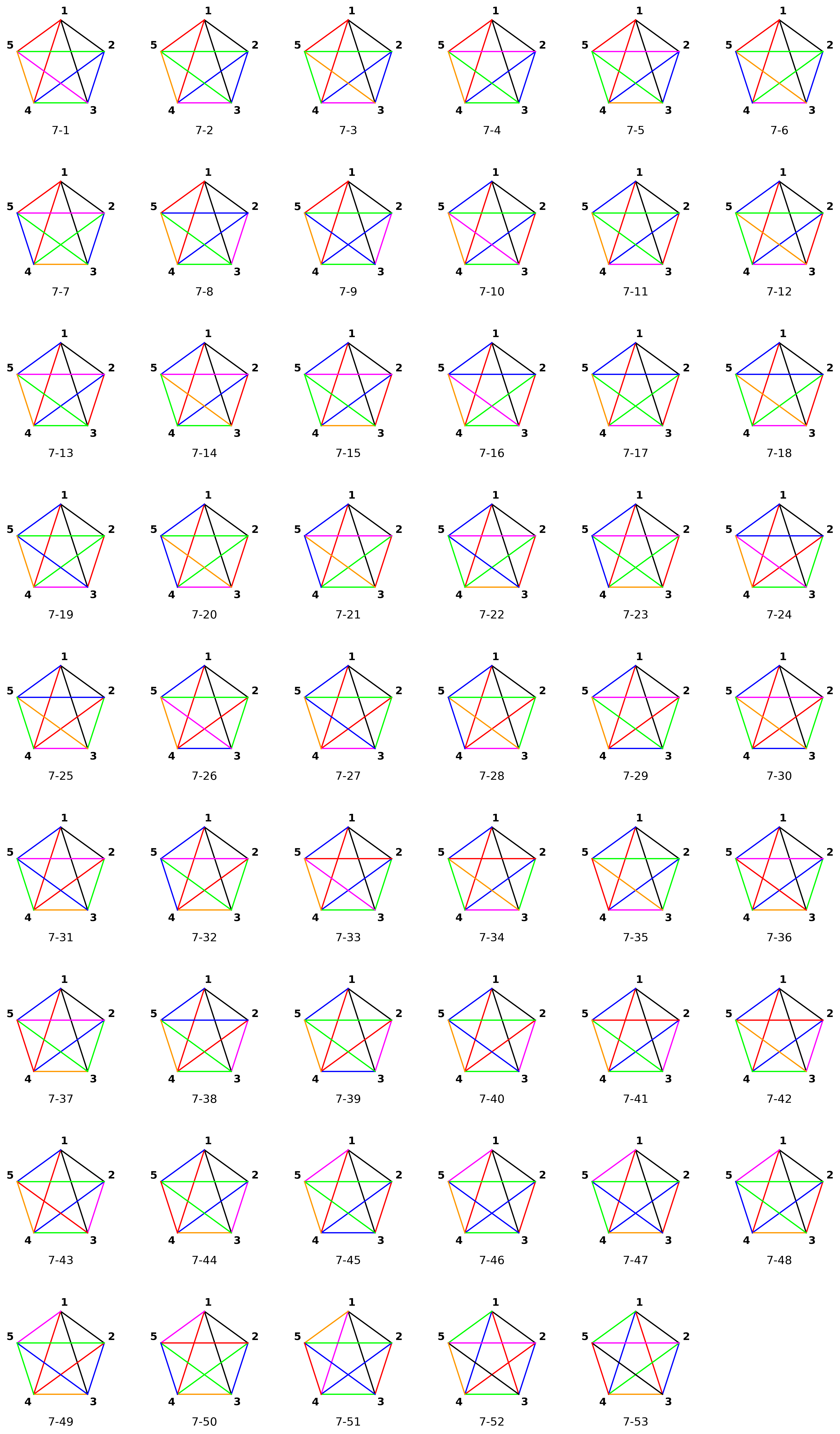}
			\caption{The feasible graphs of five six‑partite OPSs with the vector of the numbers of pairwise orthogonal relations $(2,2,2,2,1,1)$.}
			\label{fig:7}
		\end{figure*}

	From the perspective of structural properties, all feasible sets of five six‑partite OPSs can be classified into five categories by the vectors of the numbers of pairwise orthogonal relations, i.e.,  $(5,1,1,1,1,1)$, $(4,2,1,1,1,1)$, $(3,3,1,1,1,1)$, $(3,2,2,1,1,1)$, and $(2,2,2,2,1,1)$. For ease of illustration, orthogonality graphs are adopted to characterize the structure of each set of five six‑partite OPSs within every category. Collectively, all feasible sets of five six‑partite OPSs  correspond to  249 distinct orthogonality graphs. All feasible orthogonality graphs are enumerated in Table~\ref{tab:1} following the above classification. Note that configurations equivalent to tabulated ones under subsystem interchange or state relabeling are omitted, since they can be analyzed via identical reasoning.
	\begin{widetext}
		\begin{center}	
	\begin{table}[H]
		\centering
		\renewcommand{\arraystretch}{1.3}
		\caption{All feasible configurations of five six-partite OPSs}
		\label{tab:1}
		\begin{tabular}{@{}c@{\hspace{8pt}}c@{\hspace{8pt}}c@{\hspace{8pt}}c@{}}
			\hline\hline
			Categories
			& \shortstack{Cases can be perfectly \\ LOCC distinguished}
			& Cases for others
			& \shortstack{Total case numbers in \\ these categories} \\
			\hline
			$(5,1,1,1,1,1)$ & 6  & 0 & 6  \\
			$(4,2,1,1,1,1)$ & 37 & 0 & 37 \\
			$(3,3,1,1,1,1)$ & 27 & 0 & 27 \\
			$(3,2,2,1,1,1)$ & 125& 1 & 126\\
			$(2,2,2,2,1,1)$ & 52 & 1 & 53 \\
			\hline\hline
		\end{tabular}
	\end{table}
	\end{center}
	\end{widetext}	
	Subsequently, we present the orthogonality graphs for all feasible configurations of five six‑partite OPSs from Fig.~\ref{fig:3} to Fig.~\ref{fig:7}. For orthogonality graphs that can be perfectly distinguished by LOCC, we provide detailed proof procedures. In addition, It should be noted that we have omitted graphs which are the same as the graphs shown under interchange of parties as clearly those cases will follow the same line of reasoning.
	
	\begin{theorem}\label{thm:1}
		Five six‑partite OPSs with the vector of the numbers of pairwise orthogonal relations $(5,1,1,1,1,1)$, $(4,2,1,1,1,1)$ or $(3,3,1,1,1,1)$ can be perfectly distinguished by LOCC.
	\end{theorem}
	
	\begin{proof} Now, we present the proof by cases.
		
		\vspace{8pt}
		{\centering\large 1. Category $(5,1,1,1,1,1)$\par}
		\vspace{4pt}
		
		\begin{figure*}[tbp]
			\centering
			\includegraphics[width=0.80\linewidth]{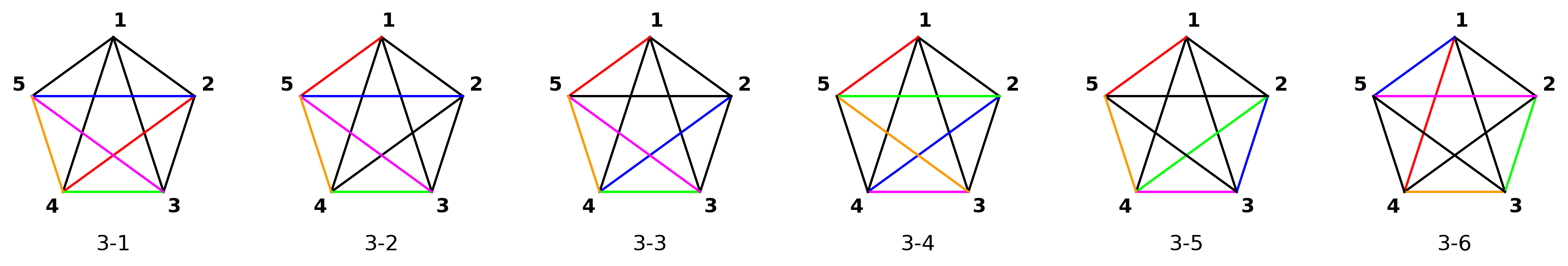}
			\caption{The feasible graphs of five six‑partite OPSs with the vector of the numbers of pairwise orthogonal relations $(5,1,1,1,1,1)$.}
			\label{fig:3}
		\end{figure*}
		
		As shown in Fig.~\ref{fig:3}. According to the orthogonality relations between state 1 and the other states, we can divide all feasible graphs into three subcases.
		
		(1) In case (3‑1), state $1$ is orthogonal to all the other states on the first subsystem. Let $\vert\alpha\rangle$ denote the normalized state of the first subsystem of state $1$. The first party Alice applies a positive operator-valued measure (POVM) described by operators $\vert\alpha\rangle\langle\alpha\vert$ and $I-\vert\alpha\rangle\langle\alpha\vert$ to the first subsystem of the measured state, where $I$ stands for the identity operator. If Alice’s measurement outcome corresponds to $\vert\alpha\rangle\langle\alpha\vert$, the measured state must be state $1$. Otherwise, if Alice’s measurement outcome corresponds to $I-\vert\alpha\rangle\langle\alpha\vert$, the measured state belongs to the set $\{2,3,4,5\}$. Obviously, states $\{2,3,4,5\}$ remain pairwise orthogonal after Alice’s measurement. By Lemma~\ref{lem:8}, states $\{2,3,4,5\}$ can be perfectly distinguished by LOCC.
		
		(2) In case (3-2), (3-3), (3-4), and (3-5), state 1 is orthogonal to each of states $\{2,3,4\}$ on the first subsystem. Obviously, state $1$ is orthogonal to state $5$ on the second subsystem, which allows the second party to distinguish between $\{1,2,3,4\}$ and $\{2,3,4,5\}$. For cases (3‑2), (3‑3), (3‑4), and (3‑5), the vector of the numbers of pairwise orthogonal relations for states $\{2,3,4,5\}$ is $(2,0,1,1,1,1)$. This implies that there exist no orthogonal relations among states $\{2,3,4,5\}$ on the second subsystem, so we only need to consider the orthogonal relations on the other five subsystems. Thus, states $\{2,3,4,5\}$ can be regarded as a set of five‑partite OPSs. By Lemma~\ref{lem:7}, states $\{2,3,4,5\}$ can be perfectly distinguished by LOCC. For cases (3‑2), the vector of the numbers of pairwise orthogonal relations for states $\{1,2,3,4\}$ is $(5,0,0,0,1,0)$. By Lemma~\ref{lem:1}, states $\{1,2,3,4\}$ can be perfectly distinguished by LOCC. For cases (3‑3), (3‑4), and (3‑5), the vectors of the numbers of pairwise orthogonal relations corresponding to the set $\{1,2,3,4\}$ are $(4,0,1,1,0,0)$, $(4,0,1,0,1,0)$, and $(3,0,1,1,1,0)$, respectively. Then for cases (3‑3), (3‑4), and (3‑5), we can prove that the set \(\{1,2,3,4\}\) in these three cases can be perfectly distinguished by LOCC, which follows from Lemma~\ref{lem:3}, Lemma~\ref{lem:3}, and Lemma~\ref{lem:6}, respectively.
		
		(3) In case (3‑6), state $1$ is orthogonal to states $5$ on the third subsystem. Therefore, based on the third subsystem, we can distinguish between the sets $\{1,2,3,4\}$ and $\{2,3,4,5\}$. For states $\{1,2,3,4\}$, the vector of the numbers of pairwise orthogonal relations is $(3,1,0,1,0,1)$. By Lemma~\ref{lem:6}, states $\{1,2,3,4\}$ can be perfectly distinguished by LOCC. For states $\{2,3,4,5\}$, the vector of the numbers of pairwise orthogonal relations is $(3,0,0,1,1,1)$. By Lemma~\ref{lem:6}, states $\{2,3,4,5\}$ can be perfectly distinguished by LOCC.
		
		In summary, five six‑partite OPSs of category $(5,1,1,1,1,1)$ can be perfectly distinguished by LOCC.

		\vspace{8pt}
		{\centering\large 2. Category $(4,2,1,1,1,1)$\par}
		\vspace{4pt}

		\begin{figure*}[tbp]
			\centering
			\includegraphics[width=0.8\linewidth]{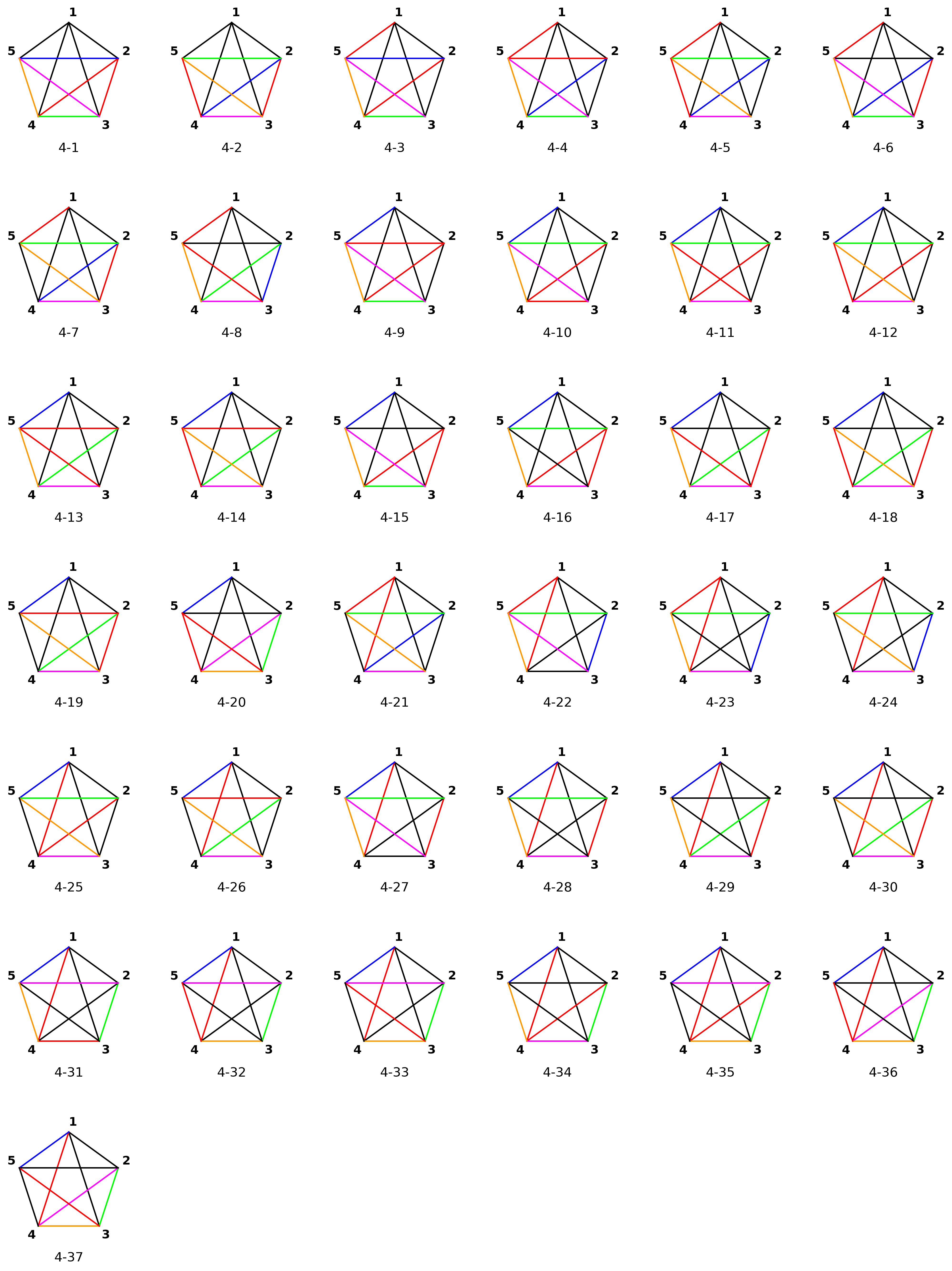}
			\caption{The feasible graphs of five six‑partite OPSs with the vector of the numbers of pairwise orthogonal relations $(4,2,1,1,1,1)$.}
			\label{fig:4}
		\end{figure*}
		
		As shown in Fig.~\ref{fig:4}. According to the orthogonality relations between state 1 and the other states, we can divide all feasible graphs into four subcases.
		
		(1) In case (4-1) and (4-2), state $1$ is orthogonal to all the other states on the first subsystem. Since the proofs are analogous, we only present the proof for case (4‑1). Based on the first subsystem, we can distinguish between the state $1$ and states $\{2,3,4,5\}$. By Lemma~\ref{lem:7}, states $\{2,3,4,5\}$ are locally distinguishable, as their vector of the numbers of pairwise orthogonal relations is $(0,2,1,1,1,1)$. The proof of case (4‑2) is similar.
		
		(2) In cases (4‑3) -- (4‑20), state $1$ is orthogonal to each of the states $\{2,3,4\}$ on the first subsystem. For case (4‑3) -- (4‑8), state $1$ is orthogonal to state $5$ on the second subsystem, but which does not allow the second party to distinguish between $\{1,2,3,4\}$ and $\{2,3,4,5\}$. Since the orthogonality satisfied by other state sets on the second subsystem, measurement may cause the second subsystem parts of the remaining states to collapse. Therefore, in the subsequent proofs, we adopt this type of orthogonality relations for discrimination: when a set of orthogonality relations is used to perform the distinction, the set must be partitioned into two subsets, and neither subset exhibits orthogonality on this subsystem. So for case (4‑3) -- (4‑8), we can first perform discrimination via the fourth subsystem. We take case (4‑3) as an example. For case (4‑3), state $3$ is orthogonal to state $4$ on the fourth subsystem, which allows the fourth party to distinguish between $\{1,2,3,5\}$ and $\{1,2,4,5\}$. For states $\{1,2,3,5\}$, the vector of the numbers of pairwise orthogonal relations is $(3,1,1,0,1,0)$. By Lemma~\ref{lem:6}, states $\{1,2,3,5\}$ can be perfectly distinguished by LOCC. For states $\{1,2,4,5\}$, the vector of the numbers of pairwise orthogonal relations is $(2,2,1,0,0,1)$. By Lemma~\ref{lem:6}, states $\{1,2,4,5\}$ can be perfectly distinguished by LOCC. Therefore, for case (4‑3), the five six‑partite OPSs can be perfectly distinguished by LOCC. The proofs for cases (4‑4) -- (4‑8) follow a similar argument. For cases (4-9) -- (4-20), state $1$ is orthogonal to state $5$ on the third subsystem, which allows the third party to distinguish between $\{1,2,3,4\}$ and $\{2,3,4,5\}$. The subsequent proof is similar to that of case (4‑3), so we do not elaborate further here.
		
		(3) In cases (4‑21) -- (4‑24), state $1$ is orthogonal to each of the states $\{4,5\}$ on the first subsystem and orthogonal to each of the states $\{4,5\}$ on the second subsystem. Similarly, only the proof for case (4‑21) is provided here. For case (4‑21), state $1$ is orthogonal to state $4$ and $5$ on the second subsystem, which allows the second party to distinguish between $\{1,2,3\}$ and $\{2,3,4,5\}$. For states $\{2,3,4,5\}$, the vector of the numbers of pairwise orthogonal relations is $(2,0,1,1,1,1)$. By Lemma~\ref{lem:7}, states $\{2,3,4,5\}$ can be perfectly distinguished by LOCC. By Lemma~\ref{lem:2}, states $\{1,2,3\}$ can be perfectly distinguished by LOCC. Therefore, for case (4‑21), the five six‑partite OPSs can be perfectly distinguished by LOCC. For cases (4‑22) -- (4‑24) the same proof method is adopted.
		
		(4) In cases (4‑25) -- (4‑37), state $1$ is orthogonal to each of the states $\{2,3\}$ on the first subsystem, to state $4$ on the second subsystem, and to state $5$ on the third subsystem. Below, we only present the proof for case (4‑25). For case (4‑25), state $1$ is orthogonal to state $5$ on the third subsystem, which allows the third party to distinguish between $\{1,2,3,4\}$ and $\{2,3,4,5\}$. For states $\{1,2,3,4\}$, the vector of the numbers of pairwise orthogonal relations is $(3,2,0,0,1,0)$. By Lemma~\ref{lem:3}, states $\{1,2,3,4\}$ can be perfectly distinguished by LOCC. For states $\{2,3,4,5\}$, the vector of the numbers of pairwise orthogonal relations is $(2,1,0,1,1,1)$. By Lemma~\ref{lem:7}, states $\{2,3,4,5\}$ can be perfectly distinguished by LOCC. Therefore, for case (4‑25), the five six‑partite OPSs can be perfectly distinguished by LOCC. Similarly, the same proof methods apply to cases (4‑26) -- (4‑37).
		
		In summary, five six‑partite OPSs of category $(4,2,1,1,1,1)$ can be perfectly distinguished by LOCC.
		
		\vspace{8pt}
		{\centering\large 3. Category $(3,3,1,1,1,1)$\par}
		\vspace{4pt}
		
		\begin{figure*}
			\centering
			\includegraphics[width=0.8\linewidth]{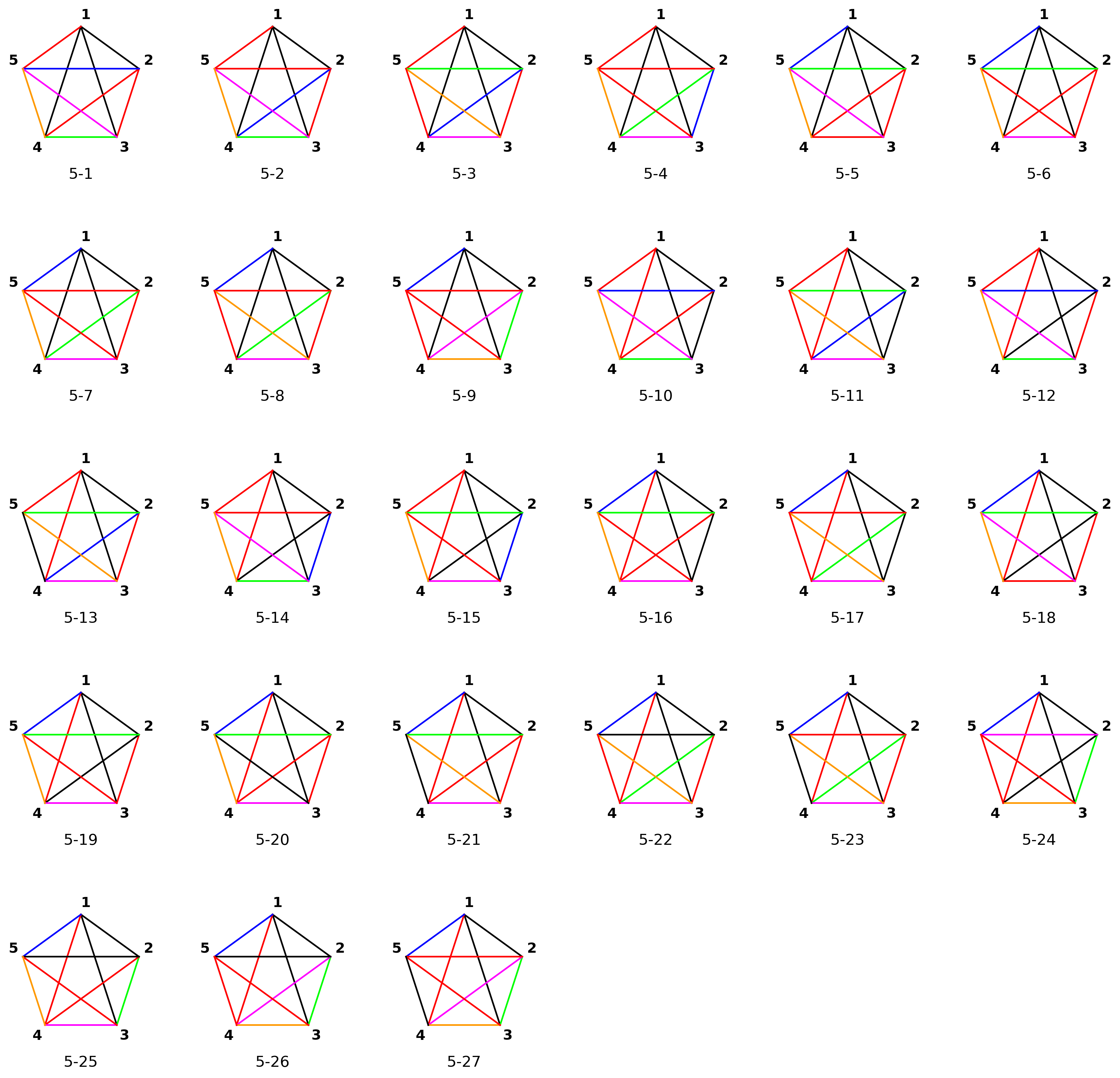}
			\caption{The feasible graphs of five six‑partite OPSs with the vector of the numbers of pairwise orthogonal relations $(3,3,1,1,1,1)$.}
			\label{fig:5}
		\end{figure*}

		As shown in Fig.~\ref{fig:5}. According to the orthogonality relations between state 1 and the other states, we can divide all feasible graphs into three subcases.
		
		(1) For cases (5‑1) -- (5‑9), state $1$ is orthogonal to each of the states $\{2,3,4\}$ on the first subsystem, which allows the first party to distinguish between $\{1,5\}$ and $\{2,3,4,5\}$. We take case (5‑1) as an example. For states \(\{1,5\}\), since they are orthogonal on the second subsystem, they can be perfectly distinguished by LOCC. For states $\{2,3,4,5\}$, the vector of the numbers of pairwise orthogonal relations is $(0,2,1,1,1,1)$. By Lemma~\ref{lem:7}, states $\{2,3,4,5\}$ can be perfectly distinguished by LOCC. Therefore, for case (5‑1), the five six‑partite OPSs can be perfectly distinguished by LOCC. Similarly, the same proof methods apply to cases (5-2) -- (5-9).
		
		(2) In cases (5‑10) -- (5‑15), state $1$ is orthogonal to each of the states $\{2,3\}$ on the first subsystem and orthogonal to each of the states $\{4,5\}$ on the second subsystem. For cases (5‑10) -- (5‑15), we can first perform discrimination via the sixth subsystem. We take case (5‑10) as an example. For case (5‑10), state $4$ is orthogonal to state $5$ on the sixth subsystem, which allows the sixth party to distinguish between $\{1,2,3,4\}$ and $\{1,2,3,5\}$. For states $\{1,2,3,4\}$, the vector of the numbers of pairwise orthogonal relations is $(3,2,0,1,0,0)$. By Lemma~\ref{lem:3}, states $\{1,2,3,4\}$ can be perfectly distinguished by LOCC. For states $\{1,2,3,5\}$, the vector of the numbers of pairwise orthogonal relations is $(3,1,1,0,1,0)$. By Lemma~\ref{lem:6}, states $\{1,2,3,5\}$ can be perfectly distinguished by LOCC. Therefore, for case (5‑10), the five six‑partite OPSs can be perfectly distinguished by LOCC. Similarly, the same proof methods apply to cases (5-11) -- (5-15).
		
		(3) In cases (5‑16) -- (5‑27), state $1$ is orthogonal to each of the states $\{2,3\}$ on the first subsystem, to state $4$ on the second subsystem, and to state $5$ on the third subsystem. For cases (5‑16) -- (5‑27), state $1$ is orthogonal to  state $5$ on the third subsystem, which allows the third party to distinguish between $\{1,2,3,4\}$ and $\{2,3,4,5\}$. We take case (5‑16) as an example. For states $\{1,2,3,4\}$, the vector of the numbers of pairwise orthogonal relations is $(3,2,0,0,1,0)$. By Lemma~\ref{lem:3}, states $\{1,2,3,4\}$ can be perfectly distinguished by LOCC. For states $\{2,3,4,5\}$, the vector of the numbers of pairwise orthogonal relations is $(1,2,0,1,1,1)$. By Lemma~\ref{lem:7}, states $\{2,3,4,5\}$ can be perfectly distinguished by LOCC. Therefore, for case (5‑16), the five six‑partite OPSs can be perfectly distinguished by LOCC. Similarly, the same proof methods apply to cases (5-17) -- (5-27).
		
		In summary, five six‑partite OPSs of category $(3,3,1,1,1,1)$ can be perfectly distinguished by LOCC. This completed the proof.\end{proof}
	
	\begin{theorem}\label{thm:2}
		Five six‑partite OPSs belonging to category $(3,2,2,1,1,1)$ or $(2,2,2,2,1,1)$ can be perfectly distinguished by LOCC, except for case (6‑73) and case (7‑51).
	\end{theorem}
	
	\begin{proof} We now proceed with the proof by cases.
		
		\vspace{8pt}
		{\centering\large 1. Category $(3,2,2,1,1,1)$\par}
		\vspace{4pt}
	
		As shown in Fig.~\ref{fig:6}. According to the orthogonality relations between state 1 and the other states, we can divide all feasible graphs into three subcases. 
		
		(1) For cases (6‑1) -- (6‑22), state $1$ is orthogonal to each of the states $\{2,3,4\}$ on the first subsystem, which allows the first party to distinguish between $\{1,5\}$ and $\{2,3,4,5\}$. We take case (6‑1) as an example. For states \(\{1,5\}\), since they are orthogonal on the second subsystem, they can be perfectly distinguished by LOCC. For states $\{2,3,4,5\}$, the vector of the numbers of pairwise orthogonal relations is $(0,1,2,1,1,1)$. By Lemma~\ref{lem:7}, states $\{2,3,4,5\}$ can be perfectly distinguished by LOCC. Therefore, for case (6‑1), the five six‑partite OPSs can be perfectly distinguished by LOCC. Similarly, the same proof methods apply to cases (6-2) -- (6-22).
		
		(2) For cases (6‑23) -- (6‑40), state $1$ is orthogonal to state $4$ and $5$ on the second subsystem, which allows the second party to distinguish between $\{2,3,4,5\}$ and $\{1,2,3\}$. We take case (6‑23) as an example. For states $\{2,3,4,5\}$, the vector of the numbers of pairwise orthogonal relations is $(1,0,2,1,1,1)$. By Lemma~\ref{lem:7}, states $\{2,3,4,5\}$ can be perfectly distinguished by LOCC. By Lemma~\ref{lem:2}, states $\{1,2,3\}$ can be perfectly distinguished by LOCC. Therefore, for case (6‑23), the five six‑partite OPSs can be perfectly distinguished by LOCC. Similarly, the same proof methods apply to cases (6‑24) -- (6‑40).
		
		(3) For cases (6‑41) -- (6‑72) and cases (6‑74) -- (6‑126), we can first perform discrimination via the sixth subsystem. We take case (6‑41) as an example. For case (6‑41), state $4$ is orthogonal to state $5$ on the sixth subsystem, which allows the sixth party to distinguish between $\{1,2,3,4\}$ and $\{1,2,3,5\}$. For states $\{1,2,3,4\}$, the vector of the numbers of pairwise orthogonal relations is $(3,2,0,1,0,0)$. By Lemma~\ref{lem:3}, states $\{1,2,3,4\}$ can be perfectly distinguished by LOCC. For states $\{1,2,3,5\}$, the vector of the numbers of pairwise orthogonal relations is $(3,0,2,0,1,0)$. By Lemma~\ref{lem:3}, states $\{1,2,3,5\}$ can be perfectly distinguished by LOCC. Therefore, for case (6‑41), the five six‑partite OPSs can be perfectly distinguished by LOCC. Similarly, the same proof methods apply to cases (6‑41) -- (6‑72) and cases (6‑74) -- (6‑126).
		
		In summary, five six‑partite OPSs of category $(3,2,2,1,1,1)$ can be perfectly distinguished by LOCC, except for case (6-73).

	\vspace{8pt}
		{\centering\large 2. Category $(2,2,2,2,1,1)$\par}
		\vspace{4pt}
        As shown in Fig.~\ref{fig:7}, according to the orthogonality relations between state 1 and the other states, we can divide all feasible graphs into three subcases. 
		
		(1) For cases (7‑1) -- (7‑9), state $1$ is orthogonal to state $2$ and $3$ on the first subsystem, which allows the first party to distinguish between $\{2,3,4,5\}$ and $\{1,4,5\}$. We take case (7-1) as an example. For states $\{2,3,4,5\}$, the vector of the numbers of pairwise orthogonal relations is $(0,0,2,2,1,1)$. By Lemma~\ref{lem:6}, states $\{2,3,4,5\}$ can be perfectly distinguished by LOCC. By Lemma~\ref{lem:2}, states $\{1,4,5\}$ can be perfectly distinguished by LOCC. Therefore, for case (7‑1), the five six‑partite OPSs can be perfectly distinguished by LOCC. Similarly, the same proof methods apply to cases (7‑2) -- (7‑9).
		
		(2) For cases (7‑10) -- (7‑50), we can first perform discrimination via the sixth subsystem. We take case (7-16) as an example. For case (7-16), state $5$ is orthogonal to state $4$ on the sixth subsystem, which allows the sixth party to distinguish between $\{1,2,3,4\}$ and $\{1,2,3,5\}$. For states $\{1,2,3,4\}$, the vector of the numbers of pairwise orthogonal relations is $(2,2,0,2,0,0)$. By graphical comparison, the graph associated with states $\{1,2,3,4\}$ reduces to graph (1-2) in Fig.~\ref{fig:1} under vertex permutation. Therefore, by Lemma~\ref{lem:4}, the states $\{1,2,3,4\}$ can be perfectly distinguished by LOCC. For states $\{1,2,3,5\}$, the vector of the numbers of pairwise orthogonal relations is $(2,1,2,0,1,0)$. By Lemma~\ref{lem:6}, states $\{1,2,3,5\}$ can be perfectly distinguished by LOCC. Therefore, for case (7‑16), the five six‑partite OPSs can be perfectly distinguished by LOCC. Similarly, the same proof methods apply to cases (7-10) -- (7-15) and cases (7-17) -- (7-50).
		
		(3) For case (7‑52) and case (7-53), state $2$ is orthogonal to state $5$ on the fifth subsystem, which allows the third party to distinguish between $\{1,2,3,4\}$ and $\{1,3,4,5\}$. We take case (7-52) as an example. For states $\{1,2,3,4\}$, the vector of the numbers of pairwise orthogonal relations is $(1,2,2,1,0,0)$. By Lemma~\ref{lem:6}, states $\{1,2,3,4\}$ can be perfectly distinguished by LOCC. For states $\{1,3,4,5\}$, the vector of the numbers of pairwise orthogonal relations is $(1,1,1,2,0,1)$. By Lemma~\ref{lem:7}, states $\{1,3,4,5\}$ can be perfectly distinguished by LOCC. Therefore, for case (7‑52), the five six‑partite OPSs can be perfectly distinguished by LOCC. Case (7-53) admits a similar proof.
		
		In summary, five six‑partite OPSs of category $(2,2,2,2,1,1)$ can be perfectly distinguished by LOCC, except for case (7‑51). This completes the proof.\end{proof}

		\begin{widetext}
			\centering
			\includegraphics[width=0.8\linewidth]{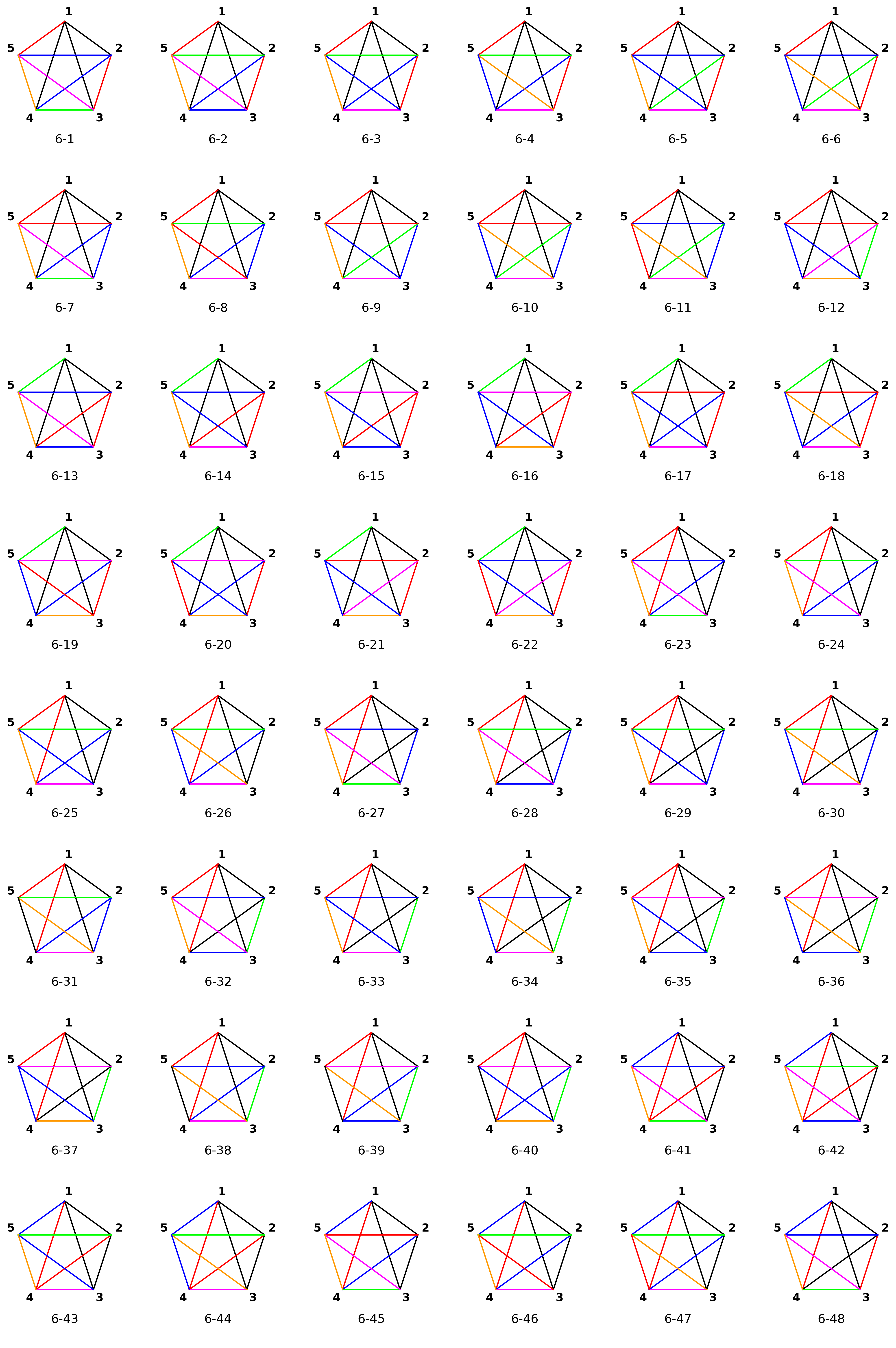}
		\end{widetext}
		\begin{widetext}
			\centering
			\includegraphics[width=0.8\linewidth]{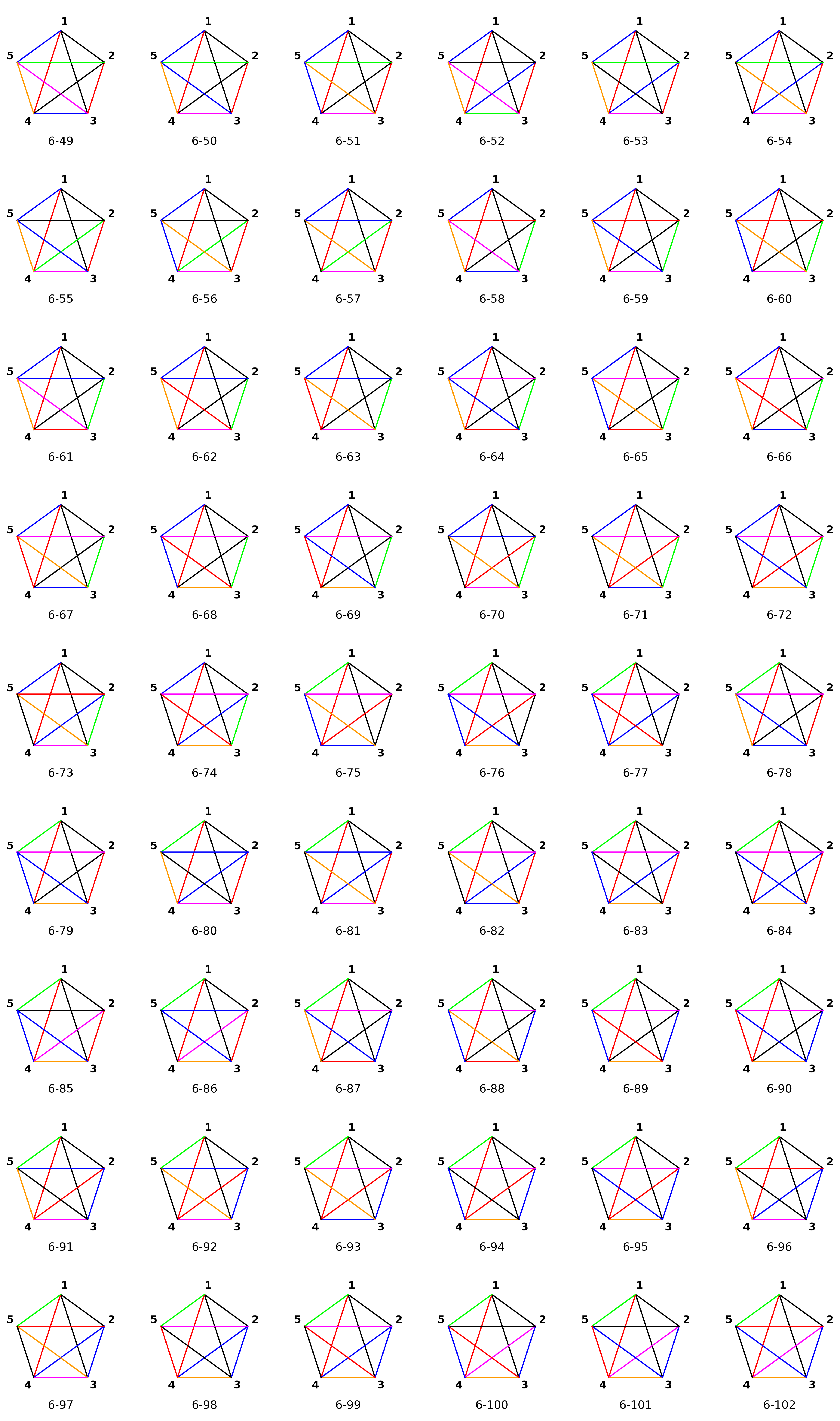}
		\end{widetext}

	\section{Analysis of Two Special Cases}
	\label{sec:4}
	In this section, we discuss the local distinguishability of two sets of five six-partite OPSs, which correspond to case (6‑73) and case (7‑51), respectively.

\begin{figure}[H]
        \centering
        \subfloat{\includegraphics[width=0.35\linewidth]{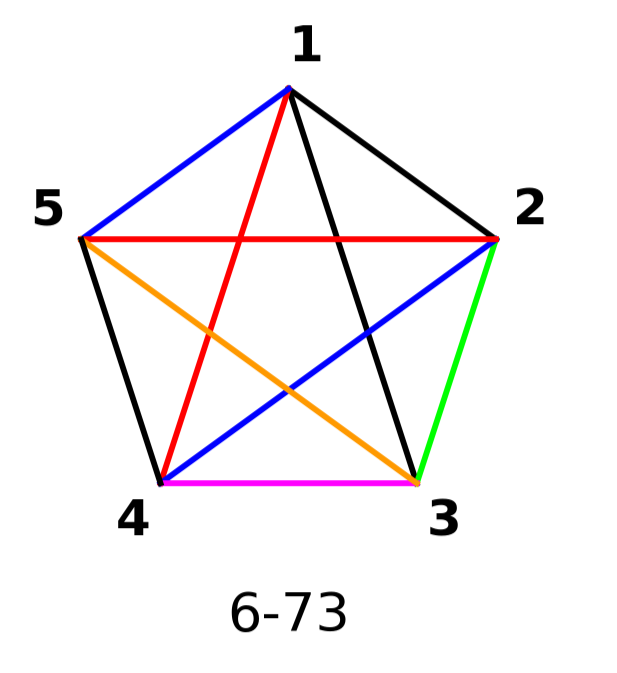}}
        \hspace{0.02\linewidth}
        \subfloat{\includegraphics[width=0.38\linewidth]{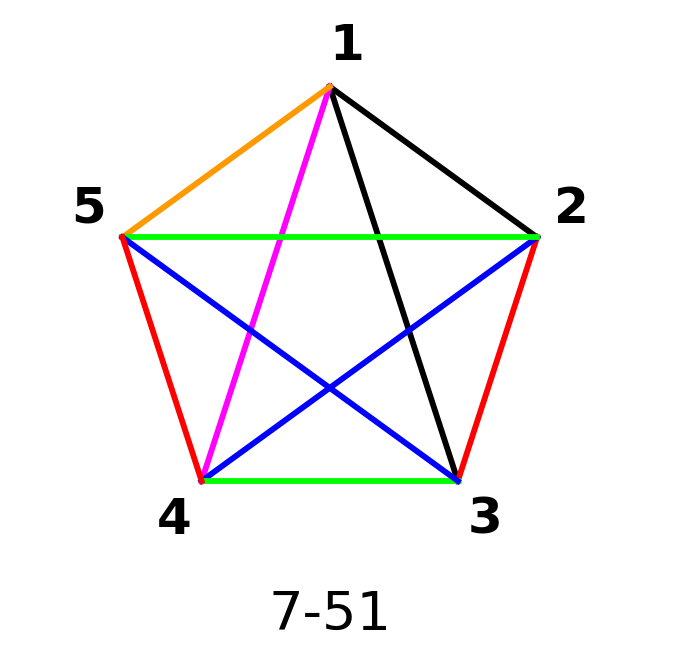}}
        \caption{Case (6‑73) and case (7‑51).}
        \label{fig:8}
      \end{figure}

	 As shown in Eq.~\eqref{eq:1}, we present a set of five six-partite OPSs that correspond to the orthogonality graph of case (6‑73) in Fig.~\ref{fig:8}. The set in 
      Eq.~\eqref{eq:1} is locally indistinguishable. We now give a proof of local indistinguishability of states in Eq.~\eqref{eq:1}.
      
      \begin{widetext}
			\begin{equation}
				\label{eq:1}
				\begin{aligned}
					|\varphi_1\rangle
					&= \frac{|0\rangle_1|0\rangle_2|0\rangle_3
					\big(|0\rangle+a_4|1\rangle\big)_4
					\big(|0\rangle+a_5|1\rangle\big)_5
					\big(|0\rangle+a_6|1\rangle\big)_6}{\sqrt{(1+|a_4|^2)(1+|a_5|^2)(1+|a_6|^2})},
					\\[1pt]
					|\varphi_2\rangle
					&= \frac{|1\rangle_1
					\big(|0\rangle+a_2|1\rangle\big)_2
					\big(|0\rangle+a_3|1\rangle\big)_3
					|0\rangle_4
					\big(|0\rangle+b_5|1\rangle\big)_5
					\big(|0\rangle+b_6|1\rangle\big)_6}{\sqrt{(1+|a_2|^2)(1+|a_3|^2)(1+|b_5|^2)(1+|b_6|^2})},
					\\[1pt]
					|\varphi_3\rangle
					&= \frac{\big(a_1|1\rangle\big)_1 
					\big(|0\rangle+b_2|1\rangle\big)_2
					\big(|0\rangle+b_3|1\rangle\big)_3
					|1\rangle_4|0\rangle_5|0\rangle_6}{\sqrt{(1+|b_2|^2)(1+|b_3|^2)|a_1|^2}},
					\\[1pt]
					|\varphi_4\rangle
					&= \frac{\big(|0\rangle-b_1|1\rangle\big)_1
					|1\rangle_2
					\big(|0\rangle-c_3|1\rangle\big)_3
					\big(|0\rangle+b_4|1\rangle\big)_4
					|1\rangle_5
					\big(|0\rangle+c_6|1\rangle\big)_6}{\sqrt{(1+|b_1|^2)(1+|c_3|^2)(1+|b_4|^2)(1+|c_6|^2})},
					\\[1pt]
					|\varphi_5\rangle
					&= \frac{\big(|0\rangle+c_1|1\rangle\big)_1
					\big(|0\rangle-c_2|1\rangle\big)_2
					|1\rangle_3
					\big(|0\rangle+c_4|1\rangle\big)_4
					\big(|0\rangle+c_5|1\rangle\big)_5
					|1\rangle_6}{\sqrt{(1+|c_1|^2)(1+|c_2|^2)(1+|c_4|^2)(1+|c_5|^2)}},
				\end{aligned}
			\end{equation}
            \end{widetext}
			where  $a_i \neq 0 $, $b_i\neq 0$ and $c_i\neq 0$ for $i\in\{1,\,2,\,\ldots,\,6\}$; $b_1^* c_1 = 1,$ $a_2^* c_2 = 1,$ $a_3^* c_3 = 1,$  $1+a_2^*b_2\neq0,$ $ 1-b_2^*c_2\neq0,$  $1+a_3^*b_3\neq0,\, 1-b_3^*c_3\neq0$; $1+a_i^*b_i\neq0$, $1+a_i^*c_i\neq0$ and $1+b_i^*c_i\neq0$ for $i\in\{4,5,6\}$.
	
		If the first party first performs an orthogonality-preserving measurement with a set of POVM elements $\{M_j^\dagger M_j\}$, where each $M_j^\dagger M_j$ can be written as
		\[
		M_j^\dagger M_j=
		\begin{pmatrix}
			m^{(j)}_{00} & m^{(j)}_{01}\\
			m^{(j)}_{10} & m^{(j)}_{11}
		\end{pmatrix}
		\]
		under the basis $\{|0\rangle, |1\rangle\}$ for $j=1,2,\dots,l$. To preserve the orthogonality of the postmeasurement states, every pair that is orthogonal only on the first subsystem must remain orthogonal after the measurement.
		
		Considering the states $|\varphi_1\rangle$ and $|\varphi_2\rangle$, we can obtain $\langle\varphi_1|\big(M_j^\dagger M_j\otimes I_{2345}\big)|\varphi_2\rangle = 0$ and $\langle\varphi_2|\big(M_j^\dagger M_j\otimes I_{2345}\big)|\varphi_1\rangle=0$. Thus, $m^{(j)}_{01}=m^{(j)}_{10}=0$. Considering the states $|\varphi_4\rangle$ and $|\varphi_5\rangle$, we can obtain $\langle\varphi_4|\big(M_j^\dagger M_j\otimes I_{2345}\big)|\varphi_5\rangle = 0$ and $\langle\varphi_5|\big(M_j^\dagger M_j\otimes I_{2345}\big)|\varphi_4\rangle=0$. Thus, we have $m^{(j)}_{00}-b_1^* c_1 m^{(j)}_{11}=0$. Since $b_1^* c_1=1$, we get $m^{(j)}_{00}=m^{(j)}_{11}$.
		
		Consequently, each POVM element performed by the first subsystem must have the form
		\[
		M_j^\dagger M_j=
		\begin{pmatrix}
			m^{(j)}_{00} & 0\\
			0 & m^{(j)}_{00}
		\end{pmatrix}.
		\]
 
   Therefore, if the measurement on the first subsystem yields outcome $j$, the conditional probability $p_1(k |j)=\langle\varphi_k|\big(M_j^\dagger M_j\otimes I_{2345}\big)|\varphi_k\rangle=m^{(j)}_{00}$ for  $k\in\{1,2,3,4,5\}$. Thus, the first party obtains the same outcome for each possible state, with equal probability, and cannot extract any useful information.
		
	The second and third parties face a situation similar to that of the first party. That is, each of the two parties is restricted to a trivial measurement if their measurement is orthogonality-preserving.
		
		For the fourth subsystem, only states 2 and 3 are orthogonal on the fourth subsystem, which allows the fourth party to distinguish between $\{1,3,4,5\}$ and $\{1,2,4,5\}$, where $i$ denotes $|\varphi_{i}\rangle$ for $i=1,\,2,\,3,\,4,\,5$. This notation is retained hereafter. For the set $\{1,3,4,5\}$, the vector of the numbers of pairwise orthogonal relations is $(2,1,1,0,1,1)$. By Lemma~\ref{lem:7}, the states $\{1,3,4,5\}$ can be perfectly distinguished by LOCC. For the set $\{1,2,4,5\}$, its orthogonality graph corresponds to graph (1-3) in Fig.~\ref{fig:1}. Therefore, by Lemma~\ref{lem:5}, the states $\{1,2,4,5\}$ cannot be perfectly distinguished by LOCC. For the fifth subsystem, only states 3 and 4 are orthogonal on the fifth subsystem, which allows the fifth party to distinguish between $\{1,2,3,5\}$ and $\{1,2,4,5\}$. For the sixth subsystem, only states 3 and 5 are orthogonal on the sixth subsystem, which allows the sixth party to distinguish between $\{1,2,3,4\}$ and $\{1,2,4,5\}$. All of them contain the set of states $\{1,2,4,5\}$. Therefore, these five states cannot be perfectly distinguished by LOCC.
		
		\begin{widetext}
			\small
			\begin{equation}
				\label{eq:2}
				\begin{aligned}
					|\varphi_1\rangle
					&= \frac{|0\rangle_1
					\big(|0\rangle+a_2|1\rangle\big)_2
					\big(|0\rangle+a_3|1\rangle\big)_3
					\big(|0\rangle+a_4|1\rangle\big)_4
					|0\rangle_5|0\rangle_6}{\sqrt{\big(1+|a_2|^2\big)\big(1+|a_3|^2\big)\big(1+|a_4|^2\big)}},
					\\[1pt]
					|\varphi_2\rangle
					&= \frac{|1\rangle_1|0\rangle_2|0\rangle_3|0\rangle_4
					\big(|0\rangle+a_5|1\rangle\big)_5
					\big(|0\rangle+a_6|1\rangle\big)_6}{\sqrt{\big(1+|a_5|^2\big)\big(1+|a_6|^2\big)}},
					\\[1pt]
					|\varphi_3\rangle
					&= \frac{\big(a_1|1\rangle\big)_1
					|1\rangle_2
					\big(|0\rangle+b_3|1\rangle\big)_3
					\big(|0\rangle+b_4|1\rangle\big)_4
					\big(|0\rangle+b_5|1\rangle\big)_5
					\big(|0\rangle+b_6|1\rangle\big)_6}{\sqrt{|a_1|^2\big(1+|b_3|^2\big)\big(1+|b_4|^2\big)\big(1+|b_5|^2\big)\big(1+|b_6|^2\big)}},
					\\[1pt]
					|\varphi_4\rangle
					&= \frac{\big(|0\rangle+b_1|1\rangle\big)_1
					\big(|0\rangle+b_2|1\rangle\big)_2
					|1\rangle_3
					\big(|0\rangle-c_4|1\rangle\big)_4
					|1\rangle_5
					\big(|0\rangle+c_6|1\rangle\big)_6}{\sqrt{\big(1+|b_1|^2\big)\big(1+|b_2|^2\big)\big(1+|c_4|^2\big)\big(1+|c_6|^2\big)}},
					\\[1pt]
					|\varphi_5\rangle
					&= \frac{\big(|0\rangle+c_1|1\rangle\big)_1
					\big(|0\rangle-c_2|1\rangle\big)_2
					\big(|0\rangle-c_3|1\rangle\big)_3
					|1\rangle_4
					\big(|0\rangle+c_5|1\rangle\big)_5
					|1\rangle_6}{\sqrt{\big(1+|c_1|^2\big)\big(1+|c_2|^2\big)\big(1+|c_3|^2\big)\big(1+|c_5|^2\big)}},
				\end{aligned}
			\end{equation}
		\end{widetext}

		where  $a_i \neq 0 $, $b_i\neq 0$ and $c_i\neq 0$ for $i\in\{1,\,2,\,\ldots,\,6\}$; $b_2^* c_2 = 1,$ $b_3^* c_3 = 1,$ $b_4^* c_4 = 1,$ $1+a_2^*b_2\neq0,$ $1-a_2^*c_2\neq0,$ $1+a_3^*b_3\neq0,$ $1-a_3^*c_3\neq0,$ $1+a_4^*b_4\neq0,$ $1-a_4^*c_4\neq0$; $1+a_i^*b_i\neq0$, $1+a_i^*c_i\neq0$ and $1+b_i^*c_i\neq0$ for $i\in\{5,6\}$.
		
		In Eq.~\eqref{eq:2}, we present a set of five six-partite OPSs in general form that correspond to case (7‑51) in Fig.~\ref{fig:8}. The set in Eq.~\eqref{eq:2} is locally indistinguishable. We now give a proof of local indistinguishability of states in Eq.~\eqref{eq:2}.
		
		Similar to the proof for the five states in Eq.~\eqref{eq:1}, for the second subsystem, the orthogonality of states 2 and 3 on the second subsystem requires the off-diagonal entries of each orthogonality-preserving POVM element to be equal to zero. For states 4 and 5, preservation of their orthogonality gives a relation between the two diagonal entries. Together with the condition $b_2^* c_2=1$, this relation implies that the two diagonal entries are equal. Hence, every such POVM element is proportional to the identity operator on the relevant local subspace, and the second subsystem cannot obtain any information for distinguishing the five states. For the third and fourth subsystems, the results are the same as that of the second subsystem; neither can obtain any information for distinguishing the five states.
		
		For the first subsystem, only state $1$ is orthogonal to state $2$ and $3$ on the first subsystem, which allows the first party to distinguish between $\{2,3,4,5\}$ and $\{1,4,5\}$. By Lemma~\ref{lem:2}, the states $\{1,4,5\}$ can be perfectly distinguished by LOCC. The set $\{2,3,4,5\}$ is structurally identical to the set $\{1,2,4,5\}$ in case (6–73), whose orthogonality graph corresponds to graph (1–3) in Fig.~\ref{fig:1}. Therefore, by Lemma~\ref{lem:5}, the states $\{2,3,4,5\}$ cannot be perfectly distinguished by LOCC. For the fifth subsystem, only states 1 and 4 are orthogonal on the fifth subsystem, which allows the fifth subsystem to distinguish between $\{1,2,3,5\}$ and $\{2,3,4,5\}$. For the sixth subsystem, only states 1 and 5 are orthogonal on the sixth subsystem, which allows the sixth subsystem to distinguish between $\{1,2,3,4\}$ and $\{2,3,4,5\}$. All of them contain the set of states $\{2,3,4,5\}$. Therefore, these five states cannot be perfectly distinguished by LOCC.
		
	\section{Conclusion}\label{sec:5}
	
	Local distinguishability of orthogonal quantum states constitutes an important research direction in quantum information. In recent years, numerous results on this topic have been reported. For instance, Ref.~\cite{ht8g-43kp}. investigates the local distinguishability of four OPSs in multipartite systems, which is followed by Ref.~\cite{53hk-wkd3}. discussing the local distinguishability of five OPSs in bipartite and tripartite systems. This raises a new question: what is the local distinguishability of five OPSs in six‑partite systems?
	
	In this paper, we investigate the local distinguishability of five six‑partite OPSs . We prove that any five six‑partite OPSs are locally distinguishable, except for two exceptional cases. Our results offer new insights into the local distinguishability of five OPSs in six‑partite systems.
\begin{acknowledgments}
	\vspace{-10pt}
    This work is supported by Natural Science Foundation of Shandong Province of China (Grant Nos. ZR2026MS1078 and ZR2023MF080) and Beijing Natural Science Foundation (Grant No. 4252014).
\end{acknowledgments}
	\bibliography{refs}
	
\end{document}